\documentclass[a4paper,11pt,final]{article}

\usepackage{amsmath,amsfonts,amssymb,bm,mathptmx,graphicx}
\usepackage[small,bf]{caption}

\DeclareFontFamily{OT1}{pzc}{}
\DeclareFontShape{OT1}{pzc}{m}{it}{<-> s*[1.125] pzcmi7t}{}
\DeclareMathAlphabet{\mathpzc}{OT1}{pzc}{m}{it}

\usepackage{hyperref}
\hypersetup{pdftitle={Type-dependent irreversible stochastic spin models for genetic regulatory networks at the level of promotion-inhibition circuitry}, pdfauthor={J. Ricardo G. Mendonça, Mário J. de Oliveira}, pdfsubject={Statistical mechanics of model systems, chemical kinetics in biological systems, Markov processes}, pdfkeywords={Genetic regulatory network, stochastic spin model, nonequilibrium statistical mechanics, repressilator, biochemical oscillations}, unicode=true, extension=pdf, pdfnewwindow=true, pdftoolbar=true, pdfmenubar=true, pdffitwindow=false, pdfstartview={FitH}, colorlinks, linkcolor=blue, citecolor=blue, urlcolor=blue}
\makeatletter
\renewcommand\@makefntext[1]{\leftskip=0.0em\hskip-0.5em\@makefnmark#1}
\makeatother
\def\leq{\leqslant}

\def\1{\mbox{\bf 1}}

\begin{document}

\pagestyle{empty}

\renewcommand{\thefootnote}{\fnsymbol{footnote}}

\begin{titlepage}

\begin{center}

{\Large {\bf Type-dependent irreversible stochastic spin models for genetic regulatory networks at the level of promotion-inhibition circuitry}}

\vspace{6ex}

{\large
{\bf J. Ricardo G. Mendon\c{c}a}$^{1,}$\footnote{Corresponding author. Email: {\tt \href{mailto:jricardo@ime.usp.br}{\nolinkurl{jricardo@usp.br}}}.}
{and} {\bf M\'{a}rio J. de Oliveira}$^{2}$
}

\vspace{1ex}

$^{1}${\it \mbox{Complex Systems Group, Escola de Artes, Ci\^{e}ncias e Humanidades, Universidade de S\~{a}o Paulo} \\ \mbox{Rua Arlindo Bettio 1000, Ermelino Matarazzo, 03828-000 S\~{a}o Paulo, SP, Brazil}}

\vspace{1ex}

$^{2}${\it \mbox{Departamento de F\'{\i}sica Geral, Instituto de F\'{\i}sica, Universidade de S\~{a}o Paulo} \\ \mbox{Caixa Postal 66318, 05314-970 S\~{a}o Paulo, SP, Brazil}}

\vspace{3ex}

{\large \bf Abstract \\}

\vspace{2ex}

\parbox{130mm}
{We describe an approach to model genetic regulatory networks at the level of promotion-in\-hi\-bi\-tion circuitry through a class of stochastic spin models that includes spatial and temporal density fluctuations in a natural way. The formalism can be viewed as an agent-based model formalism with agent behavior ruled by a classical spin-like pseudo-Hamiltonian playing the role of a local, individual objective function. A particular but otherwise generally applicable choice for the microscopic transition rates of the models also makes them of independent interest. To illustrate the formalism, we investigate (by Monte Carlo simulations) some stationary state properties of the repressilator, a synthetic three-gene network of transcriptional regulators that possesses oscillatory behavior.

\vspace{2ex}

{\noindent}{\bf Keywords}: Genetic regulatory network $\cdot$ stochastic spin model $\cdot$ nonequilibrium \linebreak statistical mechanics $\cdot$ repressilator $\cdot$ biochemical oscillations

\vspace{2ex}

{\noindent} {\bf PACS 2010}: 64.60.De $\cdot$ 82.39.-k $\cdot$ 02.50.Ga}



\end{center}

\end{titlepage}

\pagestyle{plain}


\newpage

\begin{flushright}
\small
{\it The simplest possible variable is one which takes on two values.\\ (If there is only one value, no change is possible.)} \\
S.-K. Ma, {\it Statistical Mechanics\/} (1985)
\end{flushright}


\section{\label{intro}Introduction}

Mathematical models in biology---whether in the study of ecological populations or biochemical signalling networks---are traditionally based on systems of reaction-diffusion differential equations akin to chemical kinetics ideas and techniques \cite{mathbio,lea}. The main tool of these approaches is the rate equation. If the numbers of interacting species (individuals, molecules, etc.) are sufficiently large and the system is sufficiently homogeneous (``well stirred''), the dynamics of the density profile $\bm{x}(t) = (x_{1}(t), \dots, x_{n}(t))$ of the densities $x_{i}(t)$ of each type of component can be described by the dynamical system
\begin{equation}
\label{xfg}
\frac{{\rm d}\bm{x}(t)}{{\rm d}t} = \bm{f}(\bm{x}(t))-\bm{g}(\bm{x}(t)),
\end{equation}
where $\bm{f}$ and $\bm{g}$ are limited functions denoting, respectively, the rates at which the components of the system are produced and degraded when the instantaneous density profile is $\bm{x}(t)$. Equations (\ref{xfg}) may include time-delayed terms, differential-difference terms, and stochastic perturbations as refinements
\cite{erditoth,macdon,kinetics,oleg,darren,tsimring,crowd}.

In the modelling of biochemical reaction networks, application of chemical kinetics ideas and techniques typically produce large systems of nonlinear differential equations with multiple timescales that are very difficult to solve. To circumvent these complications, and also to provide modelling tools at varied levels of abstraction, approaches based on Boolean networks, stochastic Petri nets, and rule-based formalisms, among others, have been developed \cite{bioexe,hlava,fisher,karlebach}. While some of these modelling frameworks propose innovative forms of representing biochemical reaction networks and integrating the models with laboratory tools and automation, most rely on differential equations for quantitative predictions. Chemical master equations, a mesoscopic approach to chemical kinetics based on stochastic birth and death processes, are also based on differential equations (and most of the times also on the well-stirred approximation) \cite{mcquarrie,qian}.

In this article we explore discrete state space, continous time stochastic spin models on the lattice to describe biochemical reaction and signalling networks that provide an alternative to the continuous descriptions based on rate equations. Stochastic spin models have been widely used to model interacting particle systems like exclusion and contact processes, voter models, branching and annihilating random walks, and similar models on the lattice \cite{liggett,marro}. The asymmetric, type-dependent stochastic spin models presented here were introduced in \cite{jordao} and seem promising in describing the space-time behavior of biochemical reaction and signalling networks. In particular, the fact that they deal with inhomogeneous, spatially distributed systems in a natural way provides a convenient framework to investigate the importance of space-time patterns to the efficiency of biological signalling, an important issue in the description of certain reaction cascades---e.\,g., in the immune system \cite{immune}. Here we supplement the exposition given in \cite{jordao} with a somewhat simpler notation and ``practical'' simulations of a model system aiming at an audience more interested in model building and Monte Carlo simulations.

The article is organised as follows. In section \ref{models} we describe type-dependent stochastic spin models, introduce the microscopic transition rates that model the dynamics of the promotion-inhibition circuitry and discuss the differences between the choices made for the transition rates here and the usual recipe in the context of equilibrium statistical mechanics. In this section we also remark how the formalism can be viewed as an agent-based model formalism with agent behavior ruled by a classical spin-like pseudo-Hamiltonian playing the role of a local, individual objective function. In section \ref{repressilator}, we test the formalism by means of Monte Carlo simulations of the repressilator, a three-genes genetic regulatory network of negative feedback that displays oscillatory behavior. Finally, in section \ref{summary} we summarise our results, highlight some features of the formalism presented, and indicate directions for further investigations and applications.


\section{\label{models}Type-dependent stochastic spin models}

\subsection{\label{math}Mathematical setup}

In what follows we draw heavily on \cite{jordao}, to which we refer the reader for mathematical minutiae; note, however, that our notation differs from that of \cite{jordao}. Let $\mathpzc{T} = \{a_{1}, a_{2}, \ldots, a_{n}\}$ be a finite set of $n$ types (e.\,g., molecules, genes, or proteins), $\mathpzc{S}_{a} = \{s_{a}^{(1)}, s_{a}^{(2)},$ $\ldots,$ $s_{a}^{(S_a)}\}$ the set of $S_a$ possible internal states of type $a$, and $\mathpzc{E} = \{(a,s):$ $a \in \mathpzc{T}$, $s \in \mathpzc{S}_{a}\}$. Also, let $\mathpzc{V}$ be the vertex set of a simple graph (without loops or multiple edges) of order $V=|\mathpzc{V}|$. We call the ordered pair $(i,a) \in \mathpzc{X} = \mathpzc{V} \times \mathpzc{T}$ a ``site,'' that is, an element of type $a$ lying in position $i$, and denote its internal state by $\eta_{i}^{a} \in \mathpzc{S}_{a}$. The state space of configurations $\bm{\eta}=(\eta_{i}^{a})$ is given by $\Omega = \mathpzc{S}^{\mathpzc{V}}_{a_{1}} \times \mathpzc{S}^{\mathpzc{V}}_{a_{2}} \times \cdots \times \mathpzc{S}^{\mathpzc{V}}_{a_{n}}$. Sites interact through a set of two-body interaction matrices $J_{ij}^{ab}(\,\cdot\,,\cdot\,): \mathpzc{E} \times \mathpzc{E} \rightarrow \mathbb{R}$, one for each pair of vertices $i,j \in \mathpzc{V}$. The element $J_{ij}^{ab}(\eta_{i}^{a}, \eta_{j}^{b})$ denotes the interaction strength that site $(i,a)$ in the internal state $\eta_{i}^{a}$ exerts upon site $(j,b)$ in the internal state $\eta_{j}^{b}$. Interactions between different types do not need to be symmetric, $J_{ij}^{ab} \neq J_{ij}^{ba}$; otherwise, we shall only consider isotropic interactions, $J_{ij}^{ab} = J_{ji}^{ab}$.


An example, that will be useful later, may help to clarify all these quantities. Suppose that our system is composed of three types, $A$, $B$ and $C$, so that $\mathpzc{T}=\{A,B,C\}$, and that each of these types can be in one of two states, say, inactive, that we will denote by $-1$, and active, that we will denote by $+1$, such that $\mathpzc{S}_{A}=$ $\mathpzc{S}_{B}=$ $\mathpzc{S}_{C}=$ $\{-1,+1\}$. We thus have, for each edge $(i,j)$ of a given substrate, modeled by a graph (e.\,g., a square lattice or the complete graph), an interaction strength $J_{ij}^{ab}(\eta_{i}^{a},\eta_{j}^{b})$ that can be any one of the $|\mathpzc{E} \times \mathpzc{E}| = 36$ possible combinations $J_{ij}^{AA}(-,-)$, $J_{ij}^{AA}(-,+)$, \ldots, $J_{ij}^{BC}(+,-)$, \ldots, $J_{ij}^{CC}(+,+)$.

From the matrices $J_{ij}^{ab}$ we define an ``energy'' function $H: \Omega \rightarrow \mathbb{R}$ by
\begin{equation}
\label{heta}
H(\bm{\eta}) = \sum_{(j,b) \in \mathpzc{X}}H_{j}^{b}(\bm{\eta}), \quad
H_{j}^{b}(\bm{\eta}) = \sum_{(i,a) \in \mathpzc{X}_{j}^{b}}
J_{ij}^{ab}(\eta_{i}^{a}, \eta_{j}^{b}),
\end{equation}
where $\mathpzc{X}_{j}^{b}$ is a neighborhood of $(j,b)$ that may or may not include $j$, $b$, or $(j,b)$. If $\eta_{i}^{a}$ promotes $\eta_{j}^{b}$, $J_{ij}^{ab} (\eta_{i}^{a}, \eta_{j}^{b}) < 0$, while if $\eta_{i}^{a}$ inhibits $\eta_{j}^{b}$, $J_{ij}^{ab} (\eta_{i}^{a}, \eta_{j}^{b}) > 0$. Viewed as a spin Hamiltonian, $H(\bm{\eta})$ is closely related with $n$-colour Ashkin-Teller and Potts models \cite{ashteller,potts}, but generalises them on the counts that it is in general a mixed-spins model, since the internal state spaces $\mathpzc{S}_{a}$ do not need to be identical, and that interactions between different types do not need to be symmetric.

\subsection{\label{delta}Transition rates}

Function $H(\bm{\eta})$ allows us to define a dynamics for the transitions of the internal states of the sites from the change brought by them to the value of $H(\bm{\eta})$, as with the usual stochastic spin models \cite{marro}. Here we will consider single-site transitions, although stirring can be added with some extra care. Let $\bm{\eta}_{i}^{a}(s) \in \Omega$ be the configuration given by $[\bm{\eta}_{i}^{a}(s)]_{j}^{b} = s$ if $(j,b) = (i,a)$ and $[\bm{\eta}_{i}^{a}(s)]_{j}^{b} = \eta_{j}^{b}$ otherwise. The energy cost of a transition $\bm{\eta}_{i}^{a}(r) \rightarrow \bm{\eta}_{i}^{a}(s)$ is then given by
\begin{equation}
\Delta_{i}^{a}(r,s)(\bm{\eta}) = H(\bm{\eta}_{i}^{a}(s)) - H({\bm{\eta}}_{i}^{a}(r)).
\end{equation}
Because of the asymmetry in the interactions, $\Delta_{i}^{a}(r,s)(\bm{\eta})$ decomposes into $\Delta_{i}^{a}(r,s)(\bm{\eta} \rightarrow i) + \Delta_{i}^{a}(r,s)(\bm{\eta} \leftarrow i)$, where
\begin{equation}
\label{self}
\Delta_{i}^{a}(r,s)(\bm{\eta} \rightarrow i) = \sum_{(j,b) \in \mathpzc{X}}
\Big[ J_{ji}^{ba}(\eta_{j}^{b},s) - J_{ji}^{ba}(\eta_{j}^{b},r) \Big]
\end{equation}
collects the energy difference due to the action of the sites in $\bm{\eta}$ upon the site $(i,a)$ when it flips from $\eta_{i}^{a}=r$ to $\eta_{i}^{a}=s$, and
\begin{equation}
\label{others}
\Delta_{i}^{a}(r,s)(\bm{\eta} \leftarrow i) = \sum_{(j,b) \in \mathpzc{X}}
\Big[ J_{ij}^{ab}(s, \eta_{j}^{b}) - J_{ij}^{ab}(r,\eta_{j}^{b}) \Big]
\end{equation}
collects the energy diference due to the action of the site $(i,a)$ upon the sites of $\bm{\eta}$ when it flips from $\eta_{i}^{a}=r$ to $\eta_{i}^{a}=s$. We now define a dynamics for the model specified by $H(\bm{\eta})$ through the set of single-site transitions rates
\begin{equation}
\label{rates}
c_{i}^{a}(r,s)(\bm{\eta}) = \Theta (\Delta_{i}^{a}(r,s)(\bm{\eta} \rightarrow i)),
\end{equation}
where $\Theta: \mathbb{R} \rightarrow \mathbb{R}_{+}$ is any non-increasing function obeying $\Theta(\Delta)e^{\Delta} = \Theta(-\Delta)e^{-\Delta}$.

The transition rates (\ref{rates}) depend only on the energy difference of the single site that flips, not on the global energy difference caused by the flip. From the vantage point of the flipping site, it is as if the rest of the system acted as a reservoir that goes unperturbed by the flip---only subsequent flips will eventually notice the change.  This prescription, that takes into account only the energy difference of the single site that flips, strongly resembles agent-based modeling approaches. In fact, it is as if each type in its site were an ``agent'' that analyses the situation around, evaluates its local objective function given by $H_{j}^{b}(\bm{\eta})$, and takes (or not) an action that maximises its resulting local fitness by minimizing its local objective function. The difference is that in general agent-based model rules are set by hand, and here they are provided by an energy-like functional. The same thing happens in the modeling of interacting particle systems like the contact, voter, and exclusion processes, where the transition rates are assigned without making any reference to a microscopic Hamiltonian \cite{liggett,marro}.

\subsection{\label{gibbs}A remark on the (nonequilibrium) stationary distribution}

Rule (\ref{rates}) diverts from the usual Metropolis recipe for rates used in Markov chain Monte Carlo simulations of equilibrium statistical systems, based on the global energy difference $\Delta H=H(\bm{\eta}')-H(\bm{\eta})$ between configurations, and has the important consequence that the stationary states of the model will not in general be distributed according to the Gibbs measure $\mu_{\rm G}(\bm{\eta}) \propto \exp(-H(\bm{\eta}))$, although there may be some function of $\bm{\eta}$ that renders a Gibbs-like stationary distribution for the model. For finite systems there will always be such a function, however nonlinear and nonlocal it may be; for infinite volume systems there may be none \cite{kleban,garrido}. For reversible stochastic spin models, single-site transition rates given by $c_{i}^{a}(r,s)(\bm{\eta}) = \Theta (\Delta_{i}^{a}(r,s)(\bm{\eta}))$ guarantee that the stationary state will be distributed according to $\mu_{\rm G}(\bm{\eta})$. For symmetric interactions, $J_{ij}^{ab} = J_{ij}^{ba}$, we obtain from eqs.~(\ref{self}) and (\ref{others}) that $\Delta_{i}^{a}(r,s)(\bm{\eta}) = 2\Delta_{i}^{a}(r,s)(\bm{\eta} \rightarrow i)$, and the two prescriptions coincide up to a factor of 2.

So, why should one pick the transition rates given by (\ref{rates}) instead of those that guarantee that the system will relax to its equilibrium Gibbs distribution? The answer is that the rates in (\ref{rates}) lead to forward Kolmogorov equations that, in the mean field approximation---corresponding to a well-stirred solution---and in the limit of a large number of particles are equivalent to a dynamical system $\dot{\bm{x}}(t) = \bm{V}(\bm{x}(t))$ for the density profile $\bm{x}(t) \in \mathbb{R}^{\mathpzc{S}}$, where $\mathpzc{S} = \prod_{a \in \mathpzc{T}}\mathpzc{S}_{a}$ and $\bm{V}(\bm{x}(t)): \mathbb{R}^{\mathpzc{S}} \to \mathbb{R}^{\mathpzc{S}}$ is a smooth vector field of the form $\bm{f}(\bm{x}(t))-\bm{g}(\bm{x}(t))$. The rates given by (\ref{rates}) thus allow us to establish a connection between the microscopic description in terms of the Markov jump process governed by $H(\bm{\eta})$ and macroscopic descriptions in terms of rate equations, although the rates obtained for $\bm{V}(\bm{x}(t))$ may not be related with the rates uniquely determined by the elementary chemical reactions, and the ensuing dynamical system may differ from the one obtained from the law of mass action \cite{mathbio,lea,erditoth}. This result was obtained in \cite{jordao} and is mildly related with results first obtained by T.~G.~Kurtz in the 1970s \cite{kurtz}, but the introduction of the type-dependent stochastic spin models (\ref{heta}) and the rates (\ref{rates}) is novel and provides a versatile modelling framework of independent interest.

Recent work on asymmetric Ising models \cite{godreche,mariojo} and their relationship with non-equilibrium stationary measures for stochastic evolutions and their transitions from Gibbs to non-Gibbs measures and vice-versa through dynamic bifurcations \cite{nongibbs} may become of importance in the undestanding of the non-equilibrium stationary state properties of type-dependent stochastic spin models and related models.



\section{\label{repressilator}A type-dependent stochastic spin model for the repressilator}

\subsection{\label{ising}The type-dependent stochastic Ising model}

The simplest type-dependent stochastic spin model has all internal state spaces $\mathpzc{S}_{a} = \{-1, +1\}$ and will be referred to as type-dependent stochastic Ising model (TDSIM). The most general two-body interaction $H_{j}^{b}(\bm{\eta})$ for TDSIMs is, to within an irrelevant additive constant, given by
\begin{equation}
\label{tdsim}
H_{j}^{b}(\bm{\eta}) =
\sum_{(i,a) \in \mathpzc{X}_{j}^{b}} \Big[ J_{ij}^{ab}\eta_{i}^{a}\eta_{j}^{b} + A_{ij}^{ab}\eta_{i}^{a} + B_{ij}^{ab}\eta_{j}^{b} \Big],
\end{equation}
where now $J_{ij}^{ab}$, $A_{ij}^{ab}$, and $B_{ij}^{ab}$ are scalar quantities. We remark that Ising-like Hamiltonians have already been used to model gene-gene interacting networks, but within the context of equilibrium distributions \cite{wolynes}. In our dynamic approach, the rates (\ref{rates}) are as important as $H_{j}^{b}(\bm{\eta})$ itself. Note also that the present approach is only barely related with the use of Ising spins to analyse consistency and monotonicity of reaction network graphs \cite{sontag}, although the determination of $H_{j}^{b}(\bm{\eta})$ depends on such graphs.

\subsection{\label{example}The TDSIM for the repressilator}

\begin{figure}
\centering
\includegraphics[viewport=224 101 367 741, scale=0.45, angle=-90, clip]{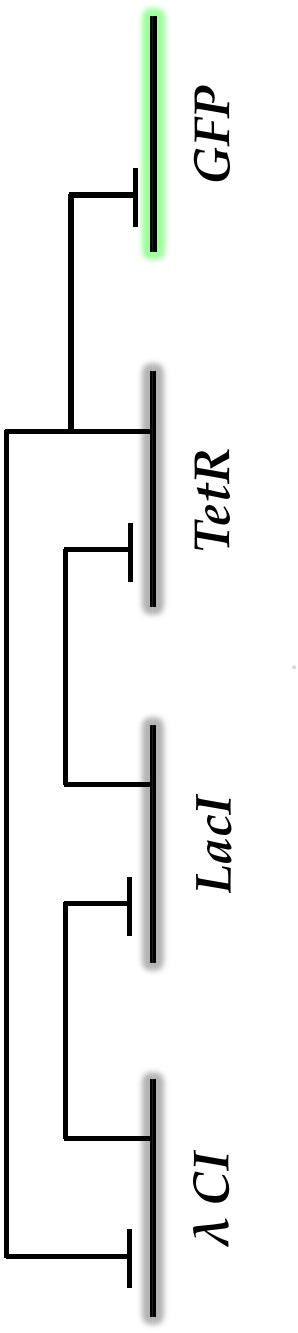}
\caption{\label{fig:repressilator}The repressilator genetic regulatory network circuit. Blunt arrows indicate inhibition through a genetic regulation mechanism briefly described in the text.}
\end{figure}

Let us illustrate the formalism by considering the repressilator, a genetic regulatory network designed to exhibit stable oscillations that are believed to be important in the determination of the circadian rythms observed in most living organisms. The repressilator was induced in the prokaryote bacteria {\it Escherichia coli\/} through a genetically engineered plasmid, together with a reporter plasmid that expresses the green fluorescent protein ({\it GFP\/}). In this system, the protein {\it LacI\/} from {\it E. coli\/} inhibits the transcription of a second gene, {\it tetR\/} from the tetracycline-resistance transposon {\it Tn10\/}, whose protein product {\it TetR\/} inhibits the transcription of a third gene, {\it cI\/} from the $\lambda$-phage, whose protein {\it CI\/} inhibits the expression of {\it lacI\/}, closing the loop of negative feedback \cite{elowitz}. This genetic regulatory network is represented in Figure~\ref{fig:repressilator}. This is clearly a highly stylised description of the true biochemical reaction network, that involves different operator sites, depends on how many proteins bind to the sites, and have lots of intermediate steps. It can, however, capture the essential nature of the interactions and is widely used to represent biochemical networks at a higher level of abstraction.

The TDSIM for the repressilator in the absence of external driving ($A_{ij}^{ab} = B_{ij}^{ab} = 0$) has three coupling constants, one for each pair of unidirectionally interacting types, all positive and that can be taken homogeneous. We take all coupling constants equal, $J^{AB} = J^{BC} = J^{CA} = J$, that despite being a considerable simplification of the full $H_{j}^{b}(\bm{\eta})$ possesses oscillatory dynamical behavior already in the mean field approximation \cite{jordao}. In this case, the two-body interaction term becomes
\begin{equation}
H_{j}(\bm{\eta}) = J \sum_{i \in \mathpzc{X}_{j}} \left[ 
\eta_{i}^{A}\eta_{j}^{B} + \eta_{i}^{B}\eta_{j}^{C} + \eta_{i}^{C}\eta_{j}^{A} \right].
\end{equation}
The velocity vector field associated with the mean field equations for this model using a heat bath prescription for the rates (\ref{rates}) (cf.~below) is given by  \cite[Prop.~5.4 and Eq.~(6.3)]{jordao}
\begin{equation}
V(x_{a}) = e^{-Jx_{b}}-\big(e^{Jx_{b}}+e^{-Jx_{b}}\big)x_{a},
\end{equation}
where $x_{a}=x_{a}^{+}(t)$ is the time-dependent density profile of type $a$ in state ``$+1$'' (clearly, $x_{a}^{-}(t)=1-x_{a}^{+}(t)$) and the indices $(a,b)$ run through the pairs $(A,B)$, $(B,C)$, and $(C,A)$.

In the lattice setting, the main quantities of interest are the empirical time-dependent densities
\begin{equation}
\rho_{a}^{s}(t) = \frac{1}{V} \sum_{i \in \mathpzc{V}} \delta(\eta_{i}^{a}(t),s), 
\end{equation}
where $\delta(\,\cdot\,,\cdot\,)$ is the Kronecker delta symbol. In practice, we measure $\rho_{a}(t) = (1/V) \sum_{i \in \mathpzc{V}} \eta_{i}^{a}(t)$, from which $\rho_{a}^{\pm}(t) = \frac{1}{2}(1 \pm \rho_{a}(t))$ can be easily recovered. The time evolution of these quantities in the stationary state of the model for some choices of $J$ appears in Figure~\ref{fig:rho}. All data were obtained by Monte Carlo simulations using a heat bath prescription $\Theta(\Delta) = 1/(1+e^{2\Delta})$ for the rates (\ref{rates}) in a simple square lattice of $V = 100 \times 100$ sites with periodic boundary conditions and nearest-neighbour interactions. Note that we include a given position in its own neighborhood to allow for intrasite interactions between different types. One Monte Carlo step equals $nV$ move attempts at randomly chosen sites $(i,a)$, where $n$ is the number of different types in the system.

\subsection{\label{density}Density profiles and correlation functions}

\begin{figure}
\centering
\begin{tabular}{c}
\includegraphics[viewport=240 60 420 760, scale=0.45, angle=-90]{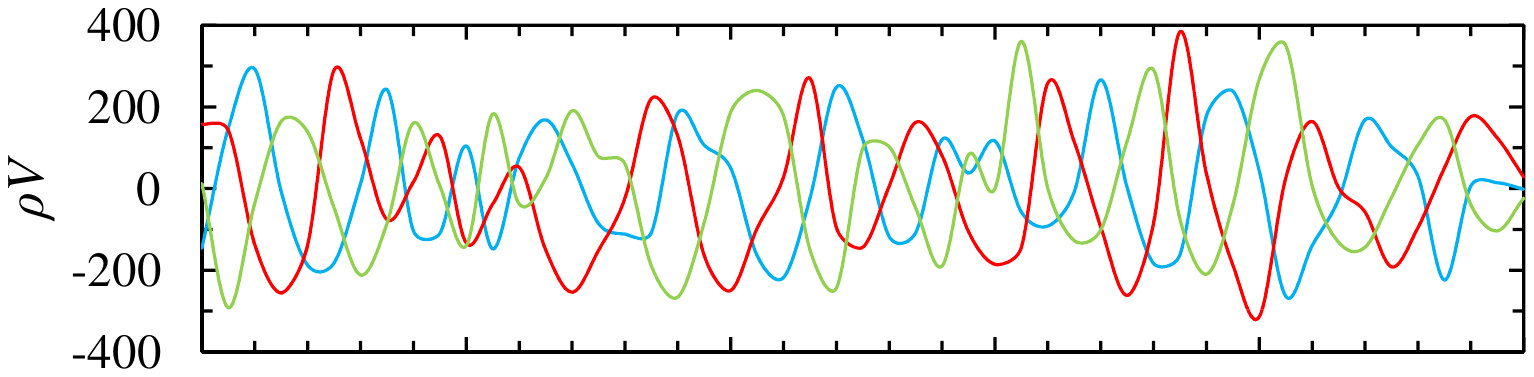} \\
\includegraphics[viewport=220 60 440 760, scale=0.45, angle=-90]{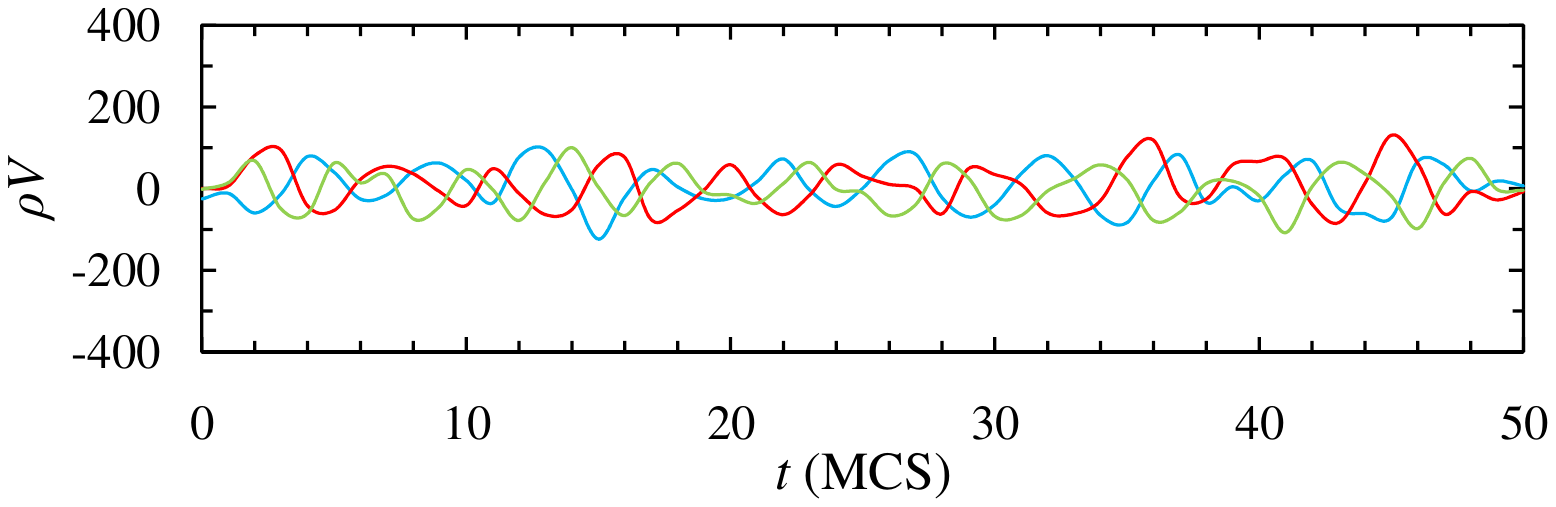}
\end{tabular}
\caption{\label{fig:rho}Evolution of the densities of the types in the stationary state of the TDSIM for the repressilator with $J=0.3$ (top panel) and $J=0.5$ (bottom panel). The densities clearly oscillate out of phase and are pairwise anticorrelated most of the time. The oscillation amplitudes at $J=0.3$ are typical in the whole range $0.07 \lesssim J \lesssim 0.42$.}
\end{figure}

Figure~\ref{fig:rho} displays the density profiles in the nonequilibrium stationary state of the model. From that figure we clearly see that the densities of different types oscillate and are out of phase. Note that the curves are mostly pairwise anticorrelated and that different types alternate in the peaks. The oscillations in figure~\ref{fig:rho} are similar to the oscillations found experimentally as well as in ODE models and stochastic simulations \cite{elowitz,comparative}. When $J \approx 0$, the types become independent or nearly independent and their densities fluctuate at will, so that we do not observe true oscillations. We could identify oscillations in our finite system for $J \gtrsim 0.07$. There is nothing special about this value, only that we can clearly observe oscillatory behavior above it. We found that the amplitudes of the oscillations vary little in the range $0.07 \lesssim J \lesssim 0.42$, but decay for $J \gtrsim 0.42$ and gets smaller as $J$ gets larger past this point. 

We found that the amplitudes of the oscillations scale like $\sqrt{V}$, signaling that the oscillations are spatially unsynchronised, since otherwise the amplitudes would scale like $V$. As a consequence, it becomes difficult to distinguish cycles or quasi-cycles out of the noise directly from the density profiles, and the analysis of correlation functions becomes preferable. This is well known from the study of population dynamics \cite{nisbet,arashiro}. We then compute the density-density time correlation functions in the stationary state,
\begin{equation}
\label{correl}
C_{ab}(t) = \lim_{T \to \infty} \frac{1}{T} \int_{0}^{T} [\rho_{a}(t+t')-\overline{\rho}_{a}][\rho_{b}(t')-\overline{\rho}_{b}] {\rm d}t',
\end{equation}
and their power spectral densities
\begin{equation}
\label{power}
S_{ab}(\omega) = \frac{1}{2\pi} \int_{-\infty}^{\infty} C_{ab}(t)e^{-i\omega t} {\rm d}t,
\end{equation}
where $\overline{\rho}_{a}$ and $\overline{\rho}_{b}$ are the average densities of types $a$ and $b$ in the stationary state. In practice, the integration limits in (\ref{correl}) and (\ref{power}) are bounded by the lengths of the time series available. In our simulations we sampled the stationary densities every $\Delta t = \frac{1}{10}$ MCS for $10^{4}$ MCS.

\begin{figure}
\centering
\begin{tabular}{c}
\includegraphics[viewport=105 51 493 792, scale=0.40, angle=-90]{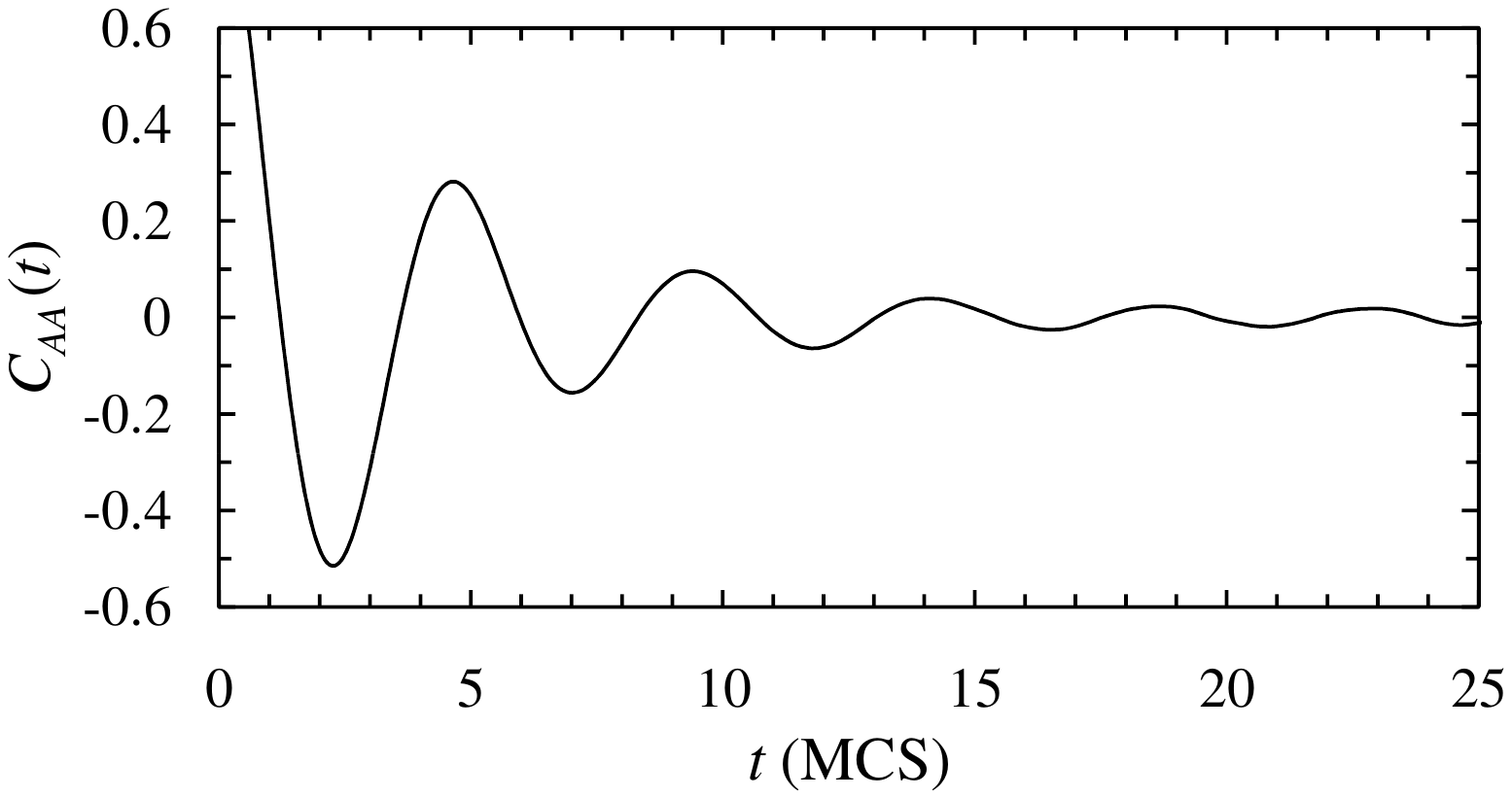} \\
\includegraphics[viewport=105 51 494 793, scale=0.40, angle=-90]{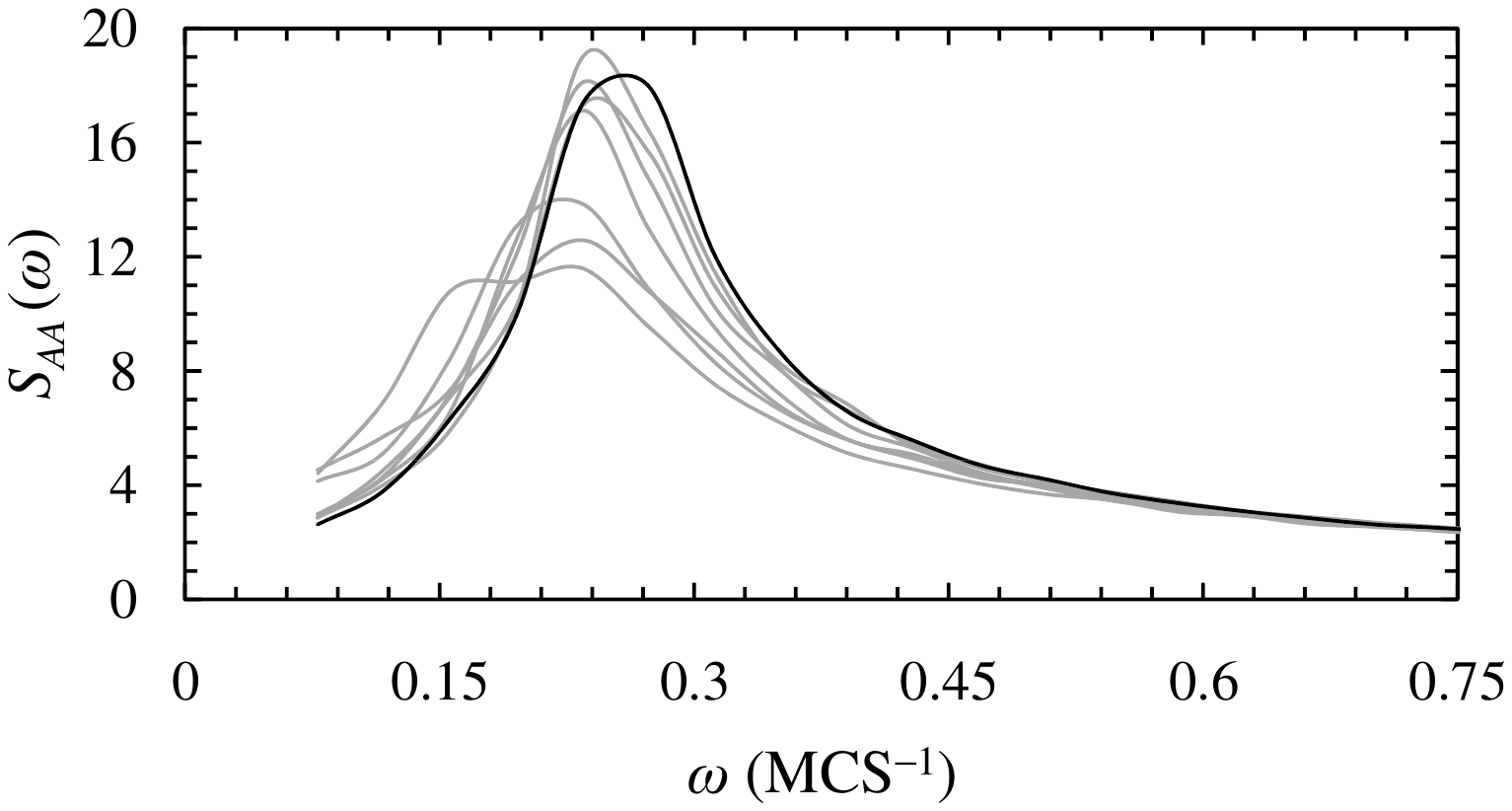}
\end{tabular}
\caption{\label{fig:fft}Autocorrelation function $C_{AA}(t)$ at $J=0.415$ normalised by its value at $t=0$ (upper panel) and some Fourier transforms $S_{AA}(\omega)$ for several different values of $0.1 \leq J \leq 0.415$ (lower panel). The curve $S_{AA}(\omega)$ for $J=0.415$ (bolder line) peaks at $\omega = 0.26 \pm 0.03$ MCS$^{-1}$.}
\end{figure}

Figure~\ref{fig:fft} displays the autocorrelation function $C_{AA}(t)$ at $J=0.415$ normalised by its value at $t=0$ and some associated Fourier transforms $S_{AA}(t)$. The other autocorrelation functions behave like $C_{AA}(t)$ because of the symmetry between the types. We see from figure~\ref{fig:fft} the decay of the autocorrelation function, typical of stochastic dynamics due to the variability of the oscillations, and the peak in $S_{AA}(\omega)$ around $\omega = 0.26 \pm 0.03$ MCS$^{-1}$ at $J=0.415$. The oscillation frequencies do not vary much with $J$ as long as $J<0.415$; otherwise, the oscillations cease almost completely for $J>0.415$.

\subsection{Onset of oscillations and the critical point}

\begin{figure}
\centering
\begin{tabular}{l@{\hspace{1em}}c@{\hspace{1em}}r}
\includegraphics[viewport=65 185 530 660, scale=0.27, angle=-90]{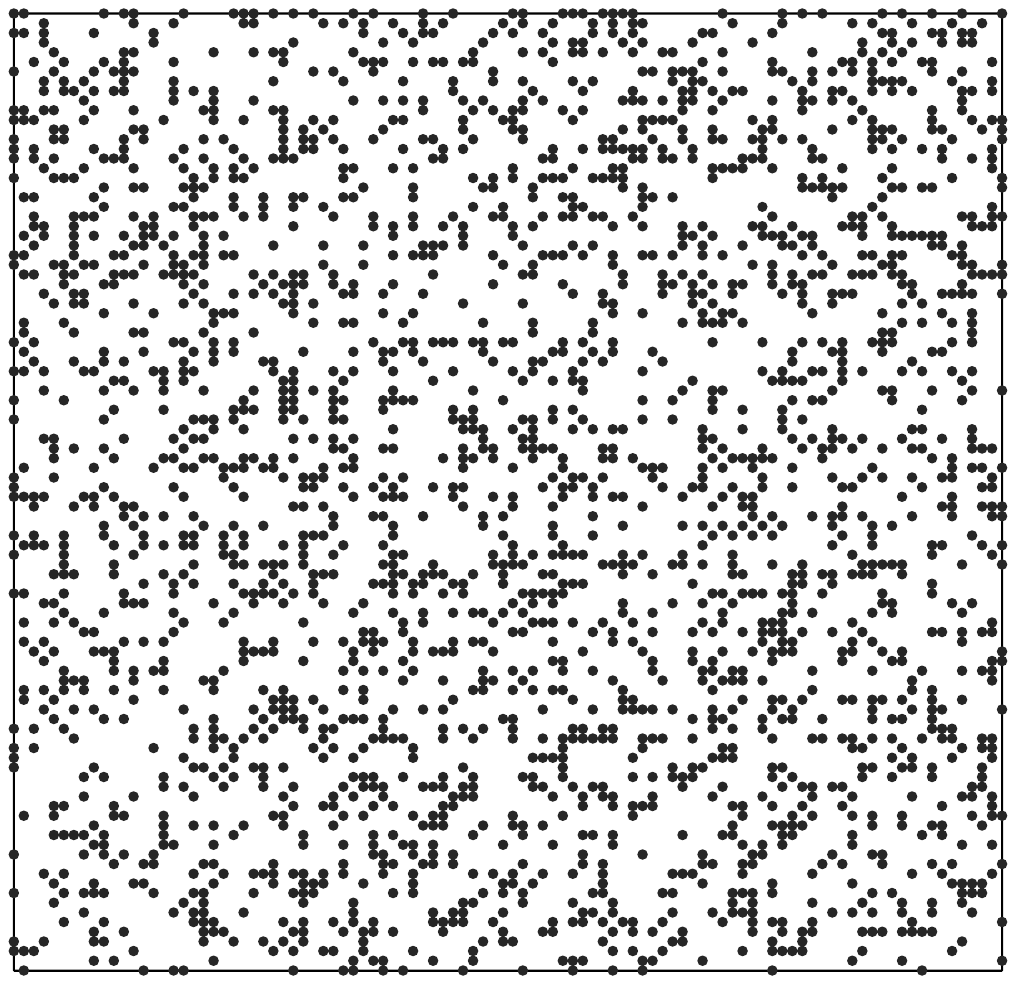} &
\includegraphics[viewport=65 185 530 660, scale=0.27, angle=-90]{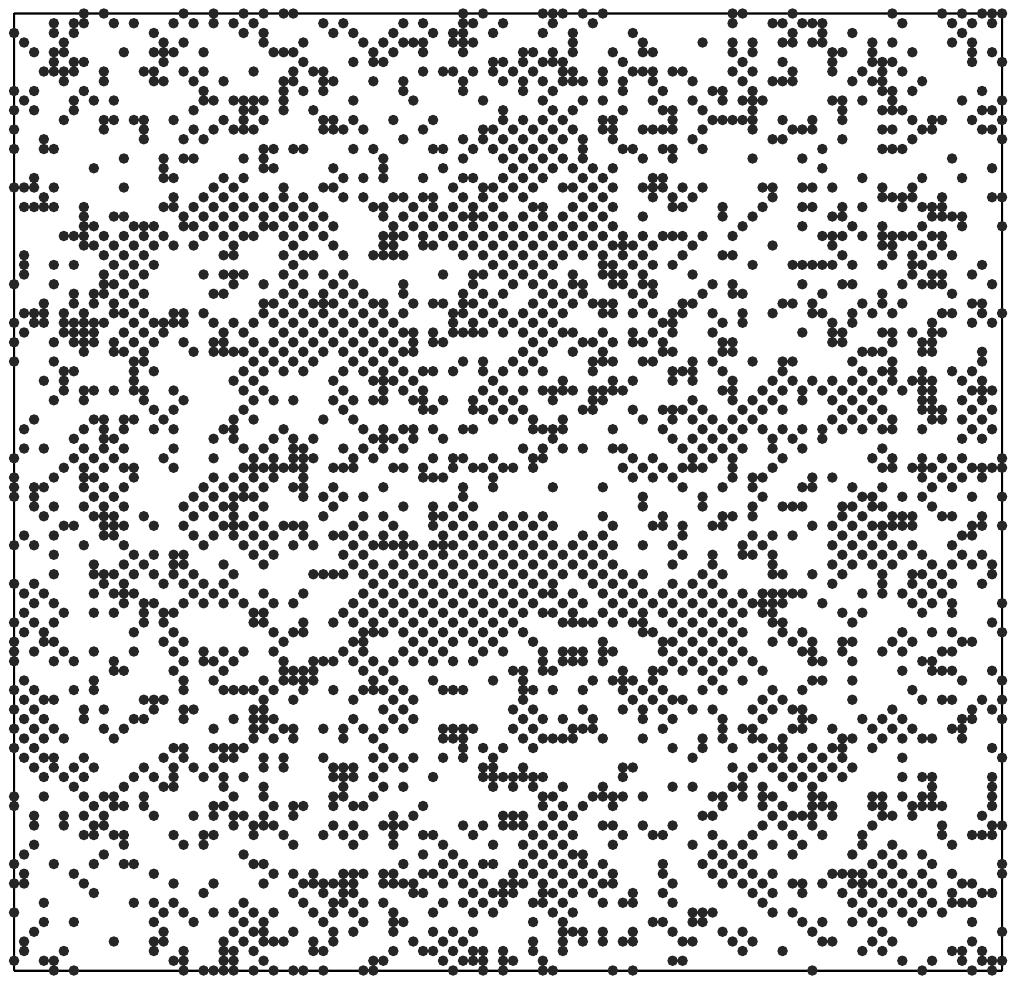} &
\includegraphics[viewport=65 185 530 660, scale=0.27, angle=-90]{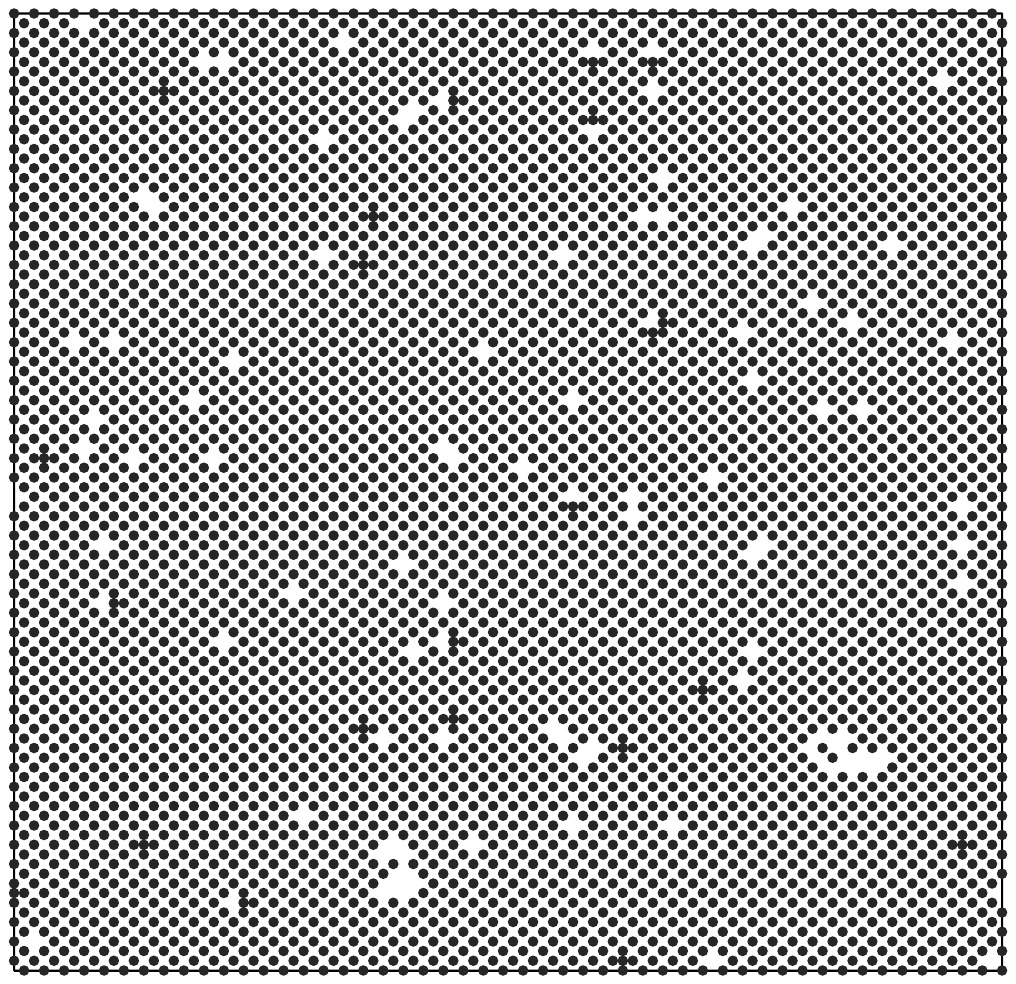}
\end{tabular}
\caption{\label{fig:abc}Correlation between the three types in a square lattice of $100 \times 100$ sites with periodic boundary conditions. The figure depicts the sites with $\eta_{i}^{a} = \eta_{i}^{b} = \eta_{i}^{c}$ (black dots) in the stationary state when $J=0.3$ (left panel), $J=0.415$ (mid panel), and $J=0.5$ (right panel). At $J=0.5$ we see an almost exact splitting into two sublattices. In this state, the remaining dynamics, responsible for the residual small amplitude oscillations shown in the botton panel of figure~\ref{fig:rho}, occurs mostly in the interstices between the sites with ``pinned'' $\eta_{i}^{a}=\eta_{i}^{b}=\eta_{i}^{c}$.}
\end{figure}

In Figure~\ref{fig:abc} we exhibit snapshots of the sites where $\eta_{i}^{a} = \eta_{i}^{b} = \eta_{i}^{c}$ in the stationary state for some values of $J$. This figure depicts a typical transition from a disordered phase to an antiferromagnetic-like phase. We clearly see how the dynamics of the types in the stationary state becomes more and more constrained by their repressors in the immediate neighborhood as $J$ gets larger, hence the smaller amplitudes in the oscillations of the densities. From figs.~\ref{fig:rho} and \ref{fig:abc} we can infer that there is a transition from a spatially uncorrelated, oscillating density stationary state to an almost frozen, non-oscillating density stationary state at $J \simeq 0.415$. We thus regard the point $J=0.415=J^{*}$ as a critical point of the model. The system does not freeze completely because of the frustration induced by the intrasite interactions between types and the form of the rates (\ref{rates}), that depend only on the single site that flips and its neighborhood, not on the state of the entire system. We located $J^{*}$ by computing the ``staggered densities'' in lattices of several sizes. 

In the dynamical mean field approximation to the same model (but with a constant external driving field independent of type) the above mentioned transition was identified with a Hopf bifurcation (when the associated real Jacobian matrix acquires a pair of pure imaginary eigenvalues) at $J^{*}=2/\cos(\pi/3)=4$ \cite{jordao}, and in a related asymmetric model with $J^{AB} = J^{BC} = J^{CA} = \delta J$ and $J^{BA} = J^{CB} = J^{AC} = (1-\delta)J$, with with $J>0$ and $0 \leq \delta \leq 1$, it was found that (with $\delta \neq \frac{1}{2}$ and again in the presence of a type-independent constant external driving) there is a Hopf bifurcation at $J^{*}=2$ \cite{navarrete}.

Our intuition about the difference in the values of $J^{*}$ observed in our simulations and in the mean field version is that the mean field version corresponds not just to a mean field version of the model, but to a continuous, off-lattice mean field version of the model. On-lattice and off-lattice versions of the same dynamics are not expected to have the same parameters; usually there are exponentials intervening in the relationship between the two limits. We remark, however, that in either case the transition at $J^{*}$ should be understood as a change in the regime of the dynamical system, not as a thermodynamic phase transition, although for systems described by a function like $H(\bm{\eta})$ the two interpretations conflate largely.


In the actual repressilator, the densities of proteins per cell oscillate with an observed period $T_{\rm obs} = 160 \pm 40$ min \cite{elowitz}. In our simulations, we found that at $J^{*}=0.415$ the period $T_{\rm sim}=3.9 \pm 0.4$ MCS. We thus have the approximate equivalence $1$ MCS~$\simeq$ $41 \pm 7$ min in the real system, and since in our simulations $1$ MCS $=100 \times 100 \times 3$ flip attempts, we can estimate that in our simulations we observed (at $J^{*}=0.415$) approximately $12 \pm 2$ transitions per second. Translation of these figures into meaningful quantities like, e.\,g., transcription and degradation rates of proteins is a delicate question that we intend to pursue elsewhere.

\section{\label{summary}Summary and perspectives}

Type-dependent irreversible interacting particle systems provide a tool to model the dynamics of biochemical reaction networks by linking influence flow diagrams like the one depicted in figure~\ref{fig:repressilator} with a model description at the same level of abstraction. The models capture the main dynamic characteristic of the system, are predictive, relatively simple, easily computable, and verifiable in a phenomenological sense. They can also be easily composed to describe interacting subsystems,
\begin{equation}
H(\bm{\eta},\bm{\xi}) = H(\bm{\eta}) + H(\bm{\xi}) + 
\sum_{(j,b)} \sum_{(i,a)} K_{ij}^{ab}(\eta_{i}^{a},\xi_{j}^{b}),
\end{equation}
in accordance with modularity principles commended by the systems approach to biology \cite{modular}.

We showed that the TDSIM for the repressilator generates density oscillations that reproduce those found experimentally and in ODE-based models. To display oscillations is a nontrivial task for nonequilibrium stationary states and is only possible for TDSIMs because the rates (\ref{rates}) do not obey the detailed balance condition with respect to its ``energy'' function (\ref{heta}).

The lattice, spatial structure of the spin systems provides a natural setting to study the spatiotemporal dynamics of extended networks, an aspect of biochemical reaction networks that has received increasing attention in the context of coupled genetic regulatory networks
\cite{signaling,shimizu,ojalvo,sgro,frustrated,mayoshi,hasty,youk} 
and also in the reconstruction of biological information flow networks from data \cite{kholodenko,kyk-ss5}. Type-dependent stochastic spin models can include diffusion through a Kawasaki-type exchange dynamics and also account for the possibility that types may be absent, not only active or inactive, in a given site, e.g., by taking some $\mathpzc{S}_{a} = \{-1,0,+1\}$. This possibility allows the modelling of deterministic and stochastic kinetics concurrently by putting on the same model types of low density (e.g., plasmid copies or enzymes) described by discrete variables $\eta_{i}^{a}$ together with types of higher density (e.g., peptides or small substrate molecules) described by an effective density in a mean-field-like description, e.g. as an external field (``pumping'') acting selectively on some types. An important feature of the formalism is that it allows for multiple occupancy of vertices. Each vertex $i \in \mathpzc{V}$ can be occupied or not by each of the possible types $a_{1}, \ldots, a_{n}$ in any of their possible internal states. Single occupancy models are usually harder to analyse than models which allow for multiple occupancy, and we believe that this is one of the strengths of the formalism.

It may be that some biochemical reaction networks give rise to TDSIMs resembling Hamiltonians known from other contexts. For example, the circadian oscillations of the proteins {\it KaiA\/}, {\it KaiB\/}, and {\it KaiC\/} in cyanobacteria can be modelled by the promotion-inhibition circuit $A \to C \dashv B \to A$ \cite{ishiura,kucho,pkj}, whose TDSIM is closely related with an Ising version of the spin-$\frac{1}{2}$ ferromagnetic-ferromagnetic-antiferromagnetic trimerised Heisenberg chain, an important model in the study of magnetisation processes in strong fields \cite{trimer}. On the other way around, the dynamics of an activator-repressor clock model that displays both toggle switch and oscillatory behaviors \cite{atkinson} may be modelled by a dimerised ferromagnetic-antiferromagnetic Ising chain that seems unexplored.


\section*{Acknowledgements}

We are indebted to Prof.~Eduardo J. Neves (IME/USP, Brazil) for having called our attention to type-dependent irreversible stochastic spin models, Manuel G. Navarrete (IME/USP, Brazil) for useful discussions, and Prof.~Anirvan M.~Sengupta (Rutgers University, USA) and Prof.~Sanjay Jain (University of Delhi, India) for having suggested potential applications to some of the ideas exposed here. This work was partially supported by CNPq, Brazil under grants PDS~151999/2010-4 (JRGM) and PQ~307407/2006-3 (MJO).


\end{document}